\documentclass[english,two column]{revtex4-1}
\usepackage[LGR,T1]{fontenc}
\usepackage[latin9]{inputenc}
\usepackage{amsmath}
\usepackage{amssymb}
\usepackage{graphicx}

\makeatletter

\DeclareRobustCommand{\greektext}{%
  \fontencoding{LGR}\selectfont\def\encodingdefault{LGR}}
\DeclareRobustCommand{\textgreek}[1]{\leavevmode{\greektext #1}}
\ProvideTextCommand{\~}{LGR}[1]{\char126#1}

\makeatother

\usepackage{babel}
\begin{document}
\title{Dynamic structure factor of one-dimensional Fermi superfluid with spin-orbit coupling}

\author{Zheng Gao$^1$}
\author{Lianyi He$^2$}
\author{Huaisong Zhao$^1$}
\email{hszhao@qdu.edu.cn}
\author{Shi-Guo Peng$^3$}
\email{pengshiguo@wipm.ac.cn}
\author{Peng Zou$^1$}
\email{phy.zoupeng@gmail.com}

\affiliation{$^1$College of Physics, Qingdao University, Qingdao 266071, China}
\affiliation{$^2$Department of Physics, Tsinghua University, Beijing 100084, China}
\affiliation{$^3$State Key Laboratory of Magnetic Resonance and Atomic and Molecular Physics, Innovation Academy for Precision Measurement Science and Technology, Chinese Academy of Sciences, Wuhan 430071, China}

\begin{abstract}
We theoretically calculate the density dynamic structure factor of
one-dimensional Fermi superfluid with Raman-type spin-orbit coupling, and analyze its main dynamical character during phase transition
between Bardeen-Cooper-Schrieffer superfluid and topological
superfluid. Our theoretical results display four kinds of single-particle excitations induced by the two-branch structure of single-particle spectrum, and the cross single-particle excitation is much easier to be seen in the spin dynamic structure factor at a small transferred momentum. Also we find a new roton-like collective mode emerges at a fixed transferred momentum $q \simeq 2k_F$, and it only appears once the system enters the topological superfluid state. The occurrence of this roton-like excitation is related to switch of global minimum in single-particle spectrum from $k=0$ to $k \simeq 2k_F$.

\end{abstract}
\maketitle

\section{Introduction}

Since the experimental realization of spin-orbit coupling (SOC) effect
in ultracold atomic gases \citep{Lin2011s,Wang2012s,Cheuk2012s},
it is possible to investigate many interesting and exotic matter states,
like stripe phase \citep{Ho2011b} and topological state \citep{Jiang2011m}, etc,
in this highly controllable system. To study properties of these many-body
matter states, lots of scattering techniques based on the interplay
between atoms and light play significant roles in enriching knowledge
about them. For example, radio frequency can often be used to study
the single-particle spectral function \citep{Chin2005r}, while two-photon
Bragg scattering technique is utilized to study both single-particle
excitations and rich collective ones \citep{Veeravalli2009b,Hoinka2017g}.

As an many-body physical quantity, dynamic structure factor
is related to the imaginary part of response function after Fourier
transformation \citep{Pitaevskii2003book}. The definition of dynamic
structure factor is related closely to a certain physical operator, which
is applied to perturb system. Usually we focus our discussion on
density operators of two spin components, which at the same time can impart a set
of momentum and energy to the system to induce a density response.
This density-related dynamic structure factor provides rich information
about the dynamics of the system. At a small transferred momentum,
the signal of dynamic structure factor is dominated by all possible
collective excitations, like Goldstone phonon excitation \citep{Hoinka2017g}, second sound \cite{Hu2022s},
Leggett excitation \citep{Leggett1966n,Zhang2017t,Zou2021d}, and
possible Higgs excitation \citep{Pekker2015a,Zhao2020d}. At a large
transferred momentum, the dynamic structure factor is mainly influenced
by the single-particle excitation \citep{Combescot2006m}, which is
determined by the many-body single-particle spectrum. Collecting all possible dynamical excitation displayed by dynamic structure
factor, we can effectively understand dynamical properties related
to a certain many-body matter state of the system.  Usually the experimental measurement of many-body physical quantities is a challenging work. However it is know that the density dynamic structure factor
is proportional to the value of centre-of-mass velocity of the system \citep{Brunello2001m}, which
makes it feasible to measure density structure factor by two-photon Bragg scattering experiment.

In this paper, we theoretically investigate one-dimensional (1D) Fermi
superfluid with Raman-type SOC effect. The system can be realized
by confining the motion of the system in other two dimension with
optical lattice technique. This system has been found to experience
a phase transition from a conventional Bardeen-Cooper-Schrieffer (BCS)
superfluid to an interesting topological superfluid by continuously
increasing an effective Zeeman field \citep{Jiang2011m,Liu2012t,Wei2012m}.
When the system comes into this topological superfluid, an impurity,
a boundary or a topological defect can generate local Majorana fermions
accompanied by a zero eigenenergy \citep{Liu2013i,Xu2014d,Liu2015s}.
Since there is no symmetry breaking during phase transition, experimentally
it is a great challenge to distinguish these two matter states. In
this paper, we theoretically calculate the density dynamic structure
factor of 1D Raman-SOC Fermi superfluid with random phase approximation \cite{AndersonPR1958}, and analysis their main dynamical
characters in both BCS and topological superfluid, which is expected
to provide some dynamical information to understand and distinguish
these two states.

This paper is organized as follows. In the next section, we will use
the language of Green\textquoteright s function to introduce the microscopic
model of 1D Fermi superfluid with Raman SOC effect, outline the mean-field
approximation and how to calculate response function with random phase
approximation. We give results of dynamic structure factor of
both BCS and topological superfluid in Sec. III. In Sec. IV and V, we give
our conclusions and acknowledgment. Some calculation details will be given in the final
appendix.

\section{Methods}

\subsection{Model and Hamiltonian}

For a two spin-components 1D Raman-SOC Fermi superfluid with \textsl{s}-wave
contact interaction, the system can be described by a model Hamiltonian 

\begin{equation}
\begin{array}{cc}
H & =\underset{\sigma}{\sum}\int dx\psi_{\sigma}^{\dagger}\left(x\right)\left[-\frac{1}{2m}\frac{\partial^{2}}{\partial x^{2}}-\mu\right]\psi_{\sigma}\left(x\right)\\
 & -h\int dx\left[\psi_{\uparrow}^{\dagger}\left(x\right)e^{i2k_{R}x}\psi_{\downarrow}\left(x\right)+H.c.\right]\\
 & +g_{1D}\int dx\psi_{\uparrow}^{\dagger}\left(x\right)\psi_{\downarrow}^{\dagger}\left(x\right)\psi_{\downarrow}\left(x\right)\psi_{\uparrow}\left(x\right),
\end{array}
\end{equation}
where $\psi_{\sigma}(\psi_{\sigma}^{\dagger})$ is the annihilation
(generation) operator of real particles with mass $m$ for spin-\textgreek{sv}
component and chemical potential $\mu$. A dimensionless parameter
$\gamma=mg_{1D}\text{/}n_{0}$ is usually used to describe the interaction
strength $g_{1D}$ of a uniform system at a bulk density $n_{0}$,
by which we can define the Fermi wave vector $k_{F}=\pi n_{0}/2$
and Fermi energy $E_{F}=k_{F}^{2}/2m$. $h$ is an effective
Zeeman field and $k_{R}$ is the recoil momentum of SOC laser beam,
both of which coming from the SOC effect. Here and in the following,
we have set $\hbar=k_{B}=1$ for simple. In many related references about SOC effect, 
a further unitary transformation will be carried out to the above Hamiltonian \cite{Liu2013i}, which
induces a term $\hat{k} \cdot \sigma$ turns out in the single-particle Hamiltonian ($\sigma$ is Pauli matrix), and the same operation also changes the physical meaning of spin index. So here we do not carry out these transformation to keep the original definition of spin index. 

A standard mean-field treatment
is done to the interaction Hamiltonian $H_{{\rm int}}=g_{1D}\int dx\psi_{\uparrow}^{\dagger}\psi_{\downarrow}^{\dagger}\psi_{\downarrow}\psi_{\uparrow}$
with the usual definition of order parameter $\Delta=-g_{1D}\left\langle \psi_{\downarrow}\psi_{\uparrow}\right\rangle $.
After Fourier transformation to the mean-field Hamiltonian,
we can obtain its expression in the momentum representation, which
reads

\begin{equation}
\begin{array}{cc}
H_{{\rm mf}} & =\underset{k\sigma}{\sum}\xi_{k}c_{k\sigma}^{\dagger}c_{k\sigma}-h\left(c_{k+k_{R}\uparrow}^{\dagger}c_{k-k_{R}\downarrow}+H.c.\right)\\
 & -\underset{k}{\sum}\left[\Delta c_{k\uparrow}^{\dagger}c_{-k\downarrow}^{\dagger}+\Delta^{*}c_{-k\downarrow}c_{k\uparrow}\right]
\end{array}
\end{equation}
with $\xi_{k}=k^{2}/2m-\mu$. Usually the order parameter $\Delta$
is a complex number. However U(1) symmetry is broken in the ground
state of the system, and the phase of $\Delta$ is pushed to randomly choose an 
constant number. Here we can just take this phase to be zero, and induce $\Delta=\Delta^{*}$.

The exact diagonalization of mean-field Hamiltonian $H_{{\rm mf}}$
is a feasible but tedious work because of the long expression of each eigenvector. Luckily this embarrassing problem can be solved by
motion equation of Green's function $\omega\left\langle \left\langle c_{1}|c_{2}\right\rangle \right\rangle =\left\langle \left[c_{1},c_{2}\right]_{+}\right\rangle +\left\langle \left\langle \left[c_{1},H_{{\rm mf}}\right]|c_{2}\right\rangle \right\rangle $,
where $c_{1}$ and $c_{2}$ are any possible fermionic operators of
the system. Finally we find that the system has six independent Green's
functions, which are 

\begin{equation}
\begin{array}{c}
G_{1}\left(k,\omega\right)\equiv\left\langle \left\langle c_{k+k_{R}\uparrow}|c_{k+k_{R}\uparrow}^{\dagger}\right\rangle \right\rangle =\underset{l}{\sum}\left[G_{1}\right]_{k}^{l}/\left(\omega-E_{k}^{l}\right),\\
G_{2}\left(k,\omega\right)\equiv\left\langle \left\langle c_{k-k_{R}\downarrow}|c_{k-k_{R}\downarrow}^{\dagger}\right\rangle \right\rangle =\underset{l}{\sum}\left[G_{2}\right]_{k}^{l}/\left(\omega-E_{k}^{l}\right),\\
\varGamma\left(k,\omega\right)\equiv\left\langle \left\langle c_{k+k_{R}\uparrow}|c_{-k-k_{R}\downarrow}\right\rangle \right\rangle =\underset{l}{\sum}\left[\varGamma\right]_{k}^{l}/\left(\omega-E_{k}^{l}\right),\\
S\left(k,\omega\right)\equiv\left\langle \left\langle c_{k-k_{R}\downarrow}|c_{k+k_{R}\uparrow}^{\dagger}\right\rangle \right\rangle =\underset{l}{\sum}\left[S\right]_{k}^{l}/\left(\omega-E_{k}^{l}\right),\\
F_{1}\left(k,\omega\right)\equiv\left\langle \left\langle c_{k+k_{R}\uparrow}|c_{-k+k_{R}\uparrow}\right\rangle \right\rangle =\underset{l}{\sum}\left[F_{1}\right]_{k}^{l}/\left(\omega-E_{k}^{l}\right),\\
F_{2}\left(k,\omega\right)\equiv\left\langle \left\langle c_{k-k_{R}\downarrow}|c_{-k-k_{R}\downarrow}\right\rangle \right\rangle =\underset{l}{\sum}\left[F_{2}\right]_{k}^{l}/\left(\omega-E_{k}^{l}\right),
\end{array}\label{eq:GF}
\end{equation}
where $l=\pm1,\pm2$ denotes respectively all four quasi-particle
energy spectrum $E_{k}^{\left(+1\right)}=-E_{k}^{\left(-1\right)}=U_{k}$
and $E_{k}^{\left(+2\right)}=-E_{k}^{\left(-2\right)}=D_{k}$. Symbols $U_{k}$
and $D_{k}$ are the up and down-branch quasi-particle
spectra, respectively,

\begin{equation}
U_{k}=\sqrt{E_{k}^{2}+h^{2}+k^{2}\lambda^{2}+2\sqrt{E_{k}^{2}h^{2}+\widetilde{\xi_{k}^{2}}k^{2}\lambda^{2}}},\label{eq:Uk}
\end{equation}

\begin{equation}
D_{k}=\sqrt{E_{k}^{2}+h^{2}+k^{2}\lambda^{2}-2\sqrt{E_{k}^{2}h^{2}+\widetilde{\xi_{k}^{2}}k^{2}\lambda^{2}}},\label{eq:Dk}
\end{equation}
with $\widetilde{\xi_{k}}=\xi_{k}+E_{R}$, $\lambda=k_{R}/m$ , $E_{R}=k_{R}^{2}/2m$
and $E_{k}=\sqrt{\widetilde{\xi_{k}^{2}}+\Delta^{2}}$. These single-particle spectra
do great influence to the static and dynamical properties of ground
state. All expressions of $\left[G_{1}\right]_{k}^{l}$,$\left[G_{2}\right]_{k}^{l}$,$\left[\Gamma\right]_{k}^{l}$,$\left[S\right]_{k}^{l}$,$\left[F_{1}\right]_{k}^{l}$
and $\left[F_{2}\right]_{k}^{l}$ will be listed in the appendix.
Based on the fluctuation and dissipation theorem, it is easy to get the relation between
 all physical quantities and corresponding Green's functions. For example,
we obtain equations of density equation

\begin{equation}
n_{1}=\sum_{k}\left\langle c_{k\uparrow}^{\dagger}c_{k\uparrow}\right\rangle =-\frac{1}{\pi}\sum_{k}\int d\omega\frac{{\rm Im}\left[G_{1}\left(k,\omega\right)\right]}{e^{\omega/T}+1},
\end{equation}

\begin{equation}
n_{2}=\sum_{k}\left\langle c_{k\downarrow}^{\dagger}c_{k\downarrow}\right\rangle =-\frac{1}{\pi}\sum_{k}\int d\omega\frac{{\rm Im}\left[G_{2}\left(k,\omega\right)\right]}{e^{\omega/T}+1},
\end{equation}
and order parameter equation
\begin{equation}
\frac{\Delta}{g_{1D}}=-\sum_{k}\left\langle c_{-k\downarrow}c_{k\uparrow}\right\rangle =\frac{1}{\pi}\sum_{k}\int d\omega\frac{{\rm Im}\left[\varGamma\left(k,\omega\right)\right]}{e^{\omega/T}+1},
\end{equation}
with Green's function $G_{1}$, $G_{2}$ and $\varGamma$ in Eq. \ref{eq:GF}
at temperature $T$. By self-consistently solving the above density
and order parameter equations, the value of chemical potential $\mu$
and order parameter $\Delta$ can be numerically calculated. 

\begin{figure}
\includegraphics[scale=0.3]{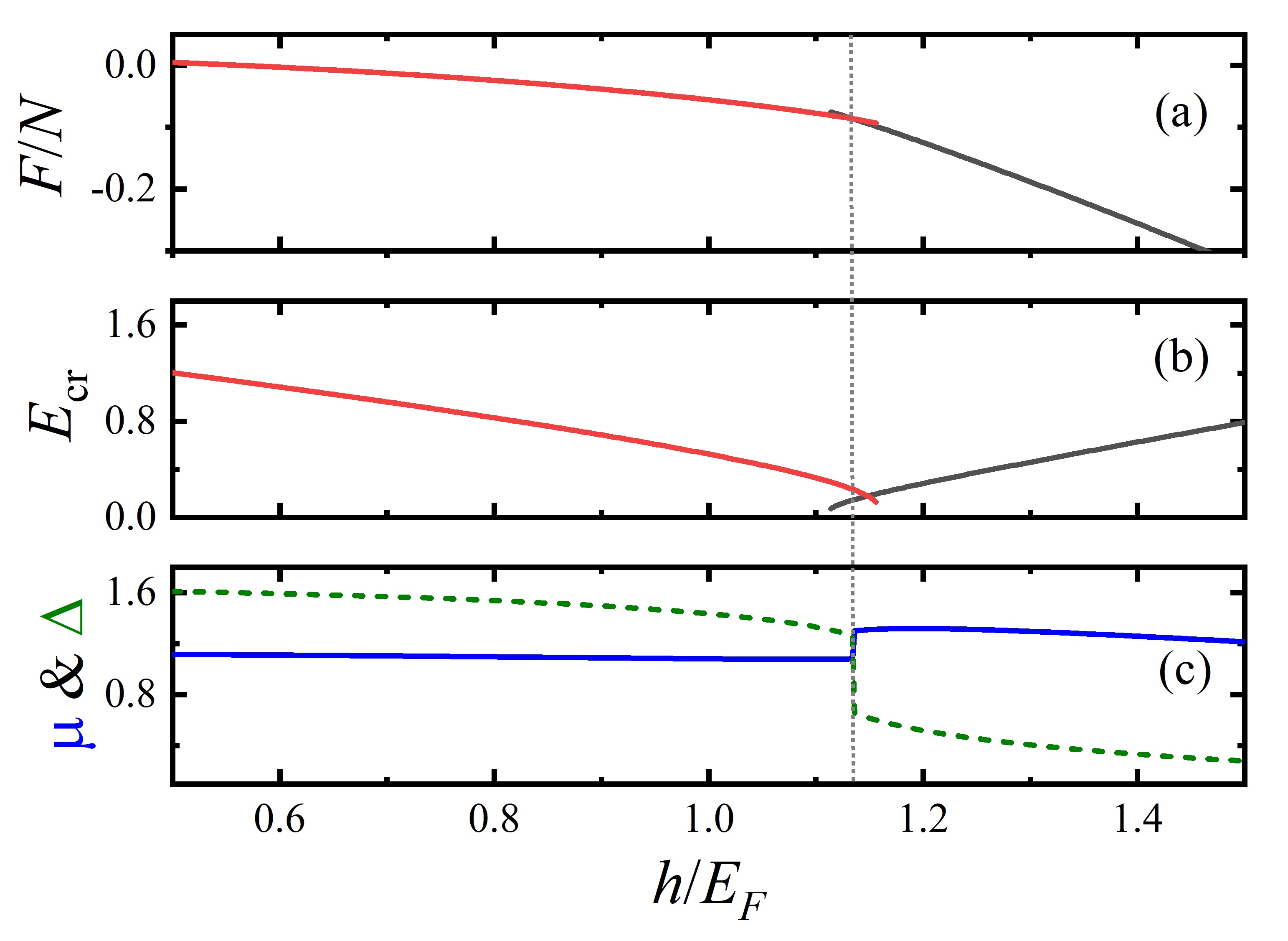}\caption{\label{Fig_transition} The distribution of Free energy in Panel (a),
$E_{{\rm cr}}=\left|h-\sqrt{\left(\mu-E_{R}\right)^{2}+\Delta^{2}}\right|$
in Panel (b), and chemical potential (blue solid line) and order parameter
(olive dashed line) at different effective Zeeman field $h$ during
the phase transition between BCS superfluid (red solid line) and topological
superfluid (black solid line). A gray dotted line marks the location of critical
value of effective Zeeman magnetic field $h_{c}=1.135E_{F}$ at $\gamma=\pi$
and $k_{R}=0.75k_{F}$.}
\end{figure}

In the following, we take an interaction strength $\gamma=\pi$ and
a typical experimental value of $k_{R}=0.75k_{F}$. As shown in Fig. \ref{Fig_transition},
the system experiences a phase transition from BCS superfluid to topological
superfluid when increasing continuously  the effective Zeeman field $h$ over a critical
value $h_{c}=1.135E_{F}$, in which the free energies of two states are equal with each other (see panel (a)). This is a first order phase transition,
during which these two states compete with each other and make chemical potential $\mu$ and order parameter
$\Delta$ experience a discontinuous variation at $h_{c}$ (see panel (c)). It should
be noticed that the critical Zeeman field $h_{c}$ here is larger
than the another critical value of Zeeman field $h$ at which the topological superfluid
just turn out and $E_{{\rm cr}}=\left|h-\sqrt{\left(\mu-E_{R}\right)^{2}+\Delta^{2}}\right|$ just touches zero (see panel (b)). 

\begin{figure}
\includegraphics[scale=0.3]{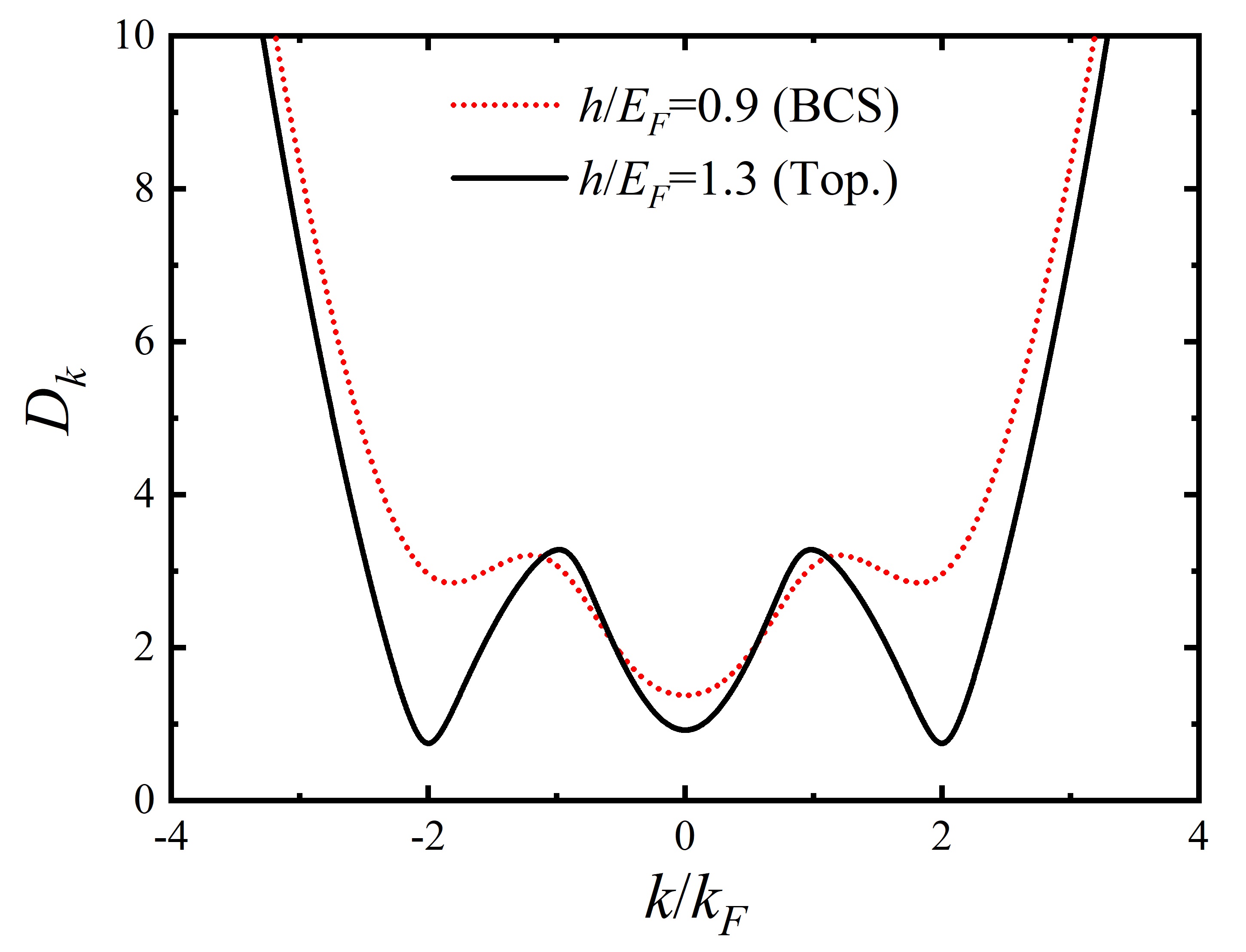}\caption{\label{Fig_uk_spectrum} The distribution of down-branch single-particle
spectrum $D_{k}$ at $\gamma=\pi$ and $k_{R}=0.75k_{F}$.}
\end{figure}

The physical origin of this phase transition can also
be understood from the geometrical structure of the down-branch single-particle spectrum $D_k$. As shown in Fig.
\ref{Fig_uk_spectrum}, the global minimum of 
$D_{k}$ experiences a switch from $k=0$ (red dotted line) to a
non-zero $k$ (black solid line), when continuously increasing Zeeman field $h$. In the critical point ($h_{c}=1.135E_{F}$), there are two options of matter state for atoms to stay in the momentum space, in which the value of chemical potential (Panel (c) of Fig. \ref{Fig_transition})
can push all atoms to stay in the regime around $k=0$, or both $k=0$
and non-zero $k$ minimum. The competition between these two situations
generate both BCS and topological superfluid with the same Free energy, and finally make the system experience this phase transition.

Next we will discuss the dynamical properties of the system and numerical methods to calculate them.

\subsection{Response function and random phase approximation}

In the Fermi superfluid, there are four different densities, which
are  denoted respectively by $n_{1}=\sum_k\left\langle c_{k\uparrow}^{\dagger}c_{k\uparrow}\right\rangle $,
$n_{2}=\sum_k\left\langle c_{-k\downarrow}^{\dagger}c_{-k\downarrow}\right\rangle $,
$n_{3}=\sum_k\left\langle c_{-k\downarrow}c_{k\uparrow}\right\rangle $ and $n_{4}=\sum_k\left\langle c_{k\uparrow}^{\dagger}c_{-k\downarrow}^{\dagger}\right\rangle $.
Due to the interaction between particles, these densities are closely
coupled with each other. Any fluctuation in each kind of density will
make the other densities generate an obvious density fluctuation
of them. This physics plays a significant role in the dynamical excitation
of the system, and also demonstrates the importance and necessity of
the term in Hamiltonian beyond mean-field theory. Random phase approximation
has been verified to be a feasible way to treat the fluctuation term
of Hamiltonian \citep{AndersonPR1958}. Comparing with experiments,
it can even obtain some quantitatively reliable predictions in three-dimensional
Fermi superfluid \citep{Zou2010q,Zou2018l}. Its prediction also qualitatively
agrees with quantum Monte Carlo data in two-dimensional Fermi system
\citep{Zhao2020d}. Here we also use the same method to
carry out calculation, to qualitatively study the dynamical excitation of 1D SOC Fermi superfluid. Its main idea is introduced in the following.

Following the standard linear response theorem, a weak external
density perturbations potential $V_{{\rm ext}}=[V_{1},V_{2},V_{3},V_{4}]$,
which carries a specific momentum $q$, is exerted to the Fermi superfluid.
The corresponding perturbation Hamiltonian is described by
$H_{{\rm ext}}=\sum_{kq}\Psi_{k+q}^{\dagger}\left(V_{1}\sigma_{1}+V_{2}\sigma_{2}+V_{3}\sigma_{3}+V_{4}\sigma_{4}\right)\Psi_{k}$.
Here $\Psi_{k}=\left[\begin{array}{cc}
c_{k\uparrow}, & c_{-k\downarrow}^{\dagger}\end{array}\right]^{T}$ is the field operator matrix in the momentum representation. Four matrices $\sigma_{1}=\left(I+\sigma_{z}\right)/2$, $\sigma_{2}=\left(I-\sigma_{z}\right)/2$, $\sigma_{3}=\left(\sigma_{x}-i\sigma_{y}\right)/2$,
and $\sigma_{4}=\left(\sigma_{x}+i\sigma_{y}\right)/2$ are defined with Pauli
matrices $\sigma_{x,y,z}$ and unit matrix $I$. This perturbation Hamiltonian $H_{\rm ext}$
will induce a density fluctuation of all densities, labeled by a matrix 
\begin{equation}
\rho_{q}=\sum_{k}\left[\begin{array}{c}
n_{kq}^{1}\\
n_{kq}^{2}\\
n_{kq}^{3}\\
n_{kq}^{4}
\end{array}\right]=\sum_{k}\left[\begin{array}{c}
\Psi_{k}^{\dagger}\sigma_{1}\Psi_{k+q}\\
\Psi_{k}^{\dagger}\sigma_{2}\Psi_{k+q}\\
\Psi_{k}^{\dagger}\sigma_{3}\Psi_{k+q}\\
\Psi_{k}^{\dagger}\sigma_{4}\Psi_{k+q}
\end{array}\right].\label{eq:fluc_den}
\end{equation}
These density fluctuation in reverse play a non-negligible role in generating a fluctuation Hamiltonian $H_{{\rm sf}}=\sum_{q}\rho_{q}^{\dagger}A_{q}$,
which is usually called self-consistent dynamical potential \citep{Liu2004c}.
Here 
\[
A_{q}=\left[\begin{array}{c}
n_{q}^{2}\\
n_{q}^{1}\\
n_{q}^{3}\\
n_{q}^{4}
\end{array}\right]=g_{1D}\sum_{k}\left[\begin{array}{c}
n_{kq}^{2}\\
n_{kq}^{1}\\
n_{kq}^{3}\\
n_{kq}^{4}
\end{array}\right]
\]
 is the strength of fluctuation potential. Different from
two or three dimension case, the contribution from $n_{q}^{3}$ and
$n_{q}^{4}$ can is not divergent and there is no need to carry on renormalization
to one dimension interaction strength $g_{1D}$. 

In a weak perturbation
situation, the amplitude of density fluctuation $\rho_{q}$ is proportional
to the external potential $V_{\rm ext}$, and they are connected to each
other by
\begin{equation}
\rho_{q}=\chi V_{{\rm ext}},\label{eq:resp}
\end{equation}
where $\chi$ is the response function of the system and includes
rich information about the dynamical excitation, however whose calculation
is usually quite hard to be carried out. As discussed above, a feasible
way to figure out this problem is to use random phase approximation, which collect effects of both $V_{{\rm ext}}$ and $V_{{\rm sf}}=M_{I}A_{q}$
to define an effective external potential 
\begin{equation}
V_{{\rm eff}}=V_{{\rm ext}}+V_{{\rm sf}}.\label{eq:veff}
\end{equation}
Then the motion of real gases in external potential $V_{{\rm ext}}$
is equivalent to the motion of mean-field gases in this effective potential
$V_{{\rm eff}}$. So the density fluctuation is connected to this effective
potential $V_{{\rm eff}}$ by
\begin{equation}
\rho_{q}=\chi^{0}V_{{\rm eff}},\label{eq:resp0}
\end{equation}
where $\chi^{0}$ is the response function in the mean-field approximation,
whose calculation is relatively much easier. Finally, with Eqs. \ref{eq:fluc_den},
\ref{eq:resp},\ref{eq:veff}, and \ref{eq:resp0}, we find $\chi$
and $\chi^{0}$ is related to each other by equation

\begin{equation}
\chi=\frac{\chi^{0}}{1-\chi^{0}M_{I}g},\label{eq:RPA}
\end{equation}
where 
\[
M_{I}=\left[\begin{array}{cccc}
0 & 1 & 0 & 0\\
1 & 0 & 0 & 0\\
0 & 0 & 0 & 1\\
0 & 0 & 1 & 0
\end{array}\right]
\]
 is a constant matrix reflecting the coupling situation of four different
densities.

Next we discuss the derivation of the mean-field response function
$\chi^{0}$, which is a $4\times4$ matrix
\begin{equation}
\chi^{0}=\left[\begin{array}{cccc}
\chi_{11}^{0} & \chi_{12}^{0} & \chi_{13}^{0} & \chi_{14}^{0}\\
\chi_{21}^{0} & \chi_{22}^{0} & \chi_{23}^{0} & \chi_{24}^{0}\\
\chi_{31}^{0} & \chi_{32}^{0} & \chi_{33}^{0} & \chi_{34}^{0}\\
\chi_{41}^{0} & \chi_{42}^{0} & \chi_{43}^{0} & \chi_{44}^{0}
\end{array}\right].\label{eq:kai0}
\end{equation}
Here any matrix element $\chi_{ij}^{0}\left(x_{1},x_{2},\tau,0\right)\equiv-\left\langle \hat{n}_{i}\left(x_{1},\tau\right)\hat{n}_{j}\left(x_{2},0\right)\right\rangle $.
In the uniform system, all response function should only be the function
of relative coordinate $x=x_{1}-x_{2}$ and imaginary time $\tau$. So a generalized
coordinate $R=\left(x,\tau\right)$ is used to go on discussing. Based
on Wick's theorem, we should consider all possible two-operators contraction
terms, which are all related to 6 independent Green's functions in Eqs. \ref{eq:GF}. We
find that the mean-field response function can be displayed by $\chi^{0}=A+B$,
in which $A$ is the mean-field response function connecting to
Green's functions $G_{1}$ , $G_{2}$ and $\Gamma$, while $B$ connecting the SOC Green's functions $S$, $F_{1}$
and $F_{2}$. For example, in the spatial and time representation,

\[
\chi_{11}^{0}\left(R\right)\equiv-\left\langle \hat{n}_{1}\left(x_{1},\tau\right)\hat{n}_{1}\left(x_{2},0\right)\right\rangle =A_{11}\left(R\right)+B_{11}\left(R\right)
\]
where $A_{11}\left(R\right)=G_{1}\left(-R\right)G_{1}\left(R\right)$
and $B_{11}\left(R\right)=F_{1}^{\dagger}\left(-R\right)F_{1}\left(R\right)$.
In the ground state ($\Delta=\Delta^{*}$), we find $F_{1}^{\dagger}=F_{1}$.
After Fourier transformation to Green's functions and with identical
relation $\frac{1}{\beta}\sum_{ip_{n}}\frac{1}{ip_{n}-\varepsilon}*\frac{1}{ip_{n}+iq_{n}-\varepsilon'}=\frac{f\left(\varepsilon\right)-f\left(\varepsilon'\right)}{iq_{n}+\varepsilon-\varepsilon'}$
($ip_{n}$ and $iq_{n}$ are Matsubara frequencies, and $f(x)$ is the Fermi distribution function), we obtain the expression
of all matrix elements in the momentum-energy representation

\begin{equation}
\chi^{0}\left(q,\omega\right)=A\left(q,\omega\right)+B\left(q,\omega\right),\label{eq:xab}
\end{equation}
where 
\[
A=\left[\begin{array}{cccc}
A_{11}, & A_{12}, & A_{13}, & A_{14}\\
A_{12}, & A_{22}, & A_{23}, & A_{24}\\
A_{14}, & A_{24}, & -A_{12}, & A_{34}\\
A_{13}, & A_{23}, & A_{43}, & -A_{12}
\end{array}\right]
\]
has 9 independent matrix elements, and 

\[
B=\left[\begin{array}{cccc}
B_{11}, & B_{12}, & B_{13}, & B_{14}\\
B_{12}, & B_{22}, & B_{23}, & B_{24}\\
B_{14}, & B_{24}, & B_{33}, & B_{34}\\
B_{13}, & B_{23}, & B_{43}, & B_{33}
\end{array}\right]
\]
has 10 independent matrix elements. All expressions of these matrix
elements are listed in the appendix.

\subsection{Dynamic structure factor}

With Eqs. \ref{eq:RPA} and \ref{eq:xab}, we get expressions
of both the density and spin response function, which are expressed
by 
\begin{equation}
\begin{array}{c}
\chi_{n}\equiv\chi_{11}+\chi_{22}+\chi_{12}+\chi_{21},\\
\chi_{s}\equiv\chi_{11}+\chi_{22}-\chi_{12}-\chi_{21}.
\end{array}
\end{equation}
And the density dynamic structure factor $S_n(q,\omega)$ and the spin one $S_s(q,\omega)$ are connected
with its corresponding response function by 

\begin{equation}
S_{n/s}=-\frac{1}{\pi}\frac{1}{1-e^{-\omega/T}}{\rm Im}\left[\chi_{n/s}\right],
\end{equation}
where $q$ and $\omega$ are the transferred momentum and energy, respectively.
The sum rules of these two dynamic structure factors have been introduced in the reference \cite{He2016d}.

\section{Results}

\begin{figure}
\includegraphics[scale=0.35]{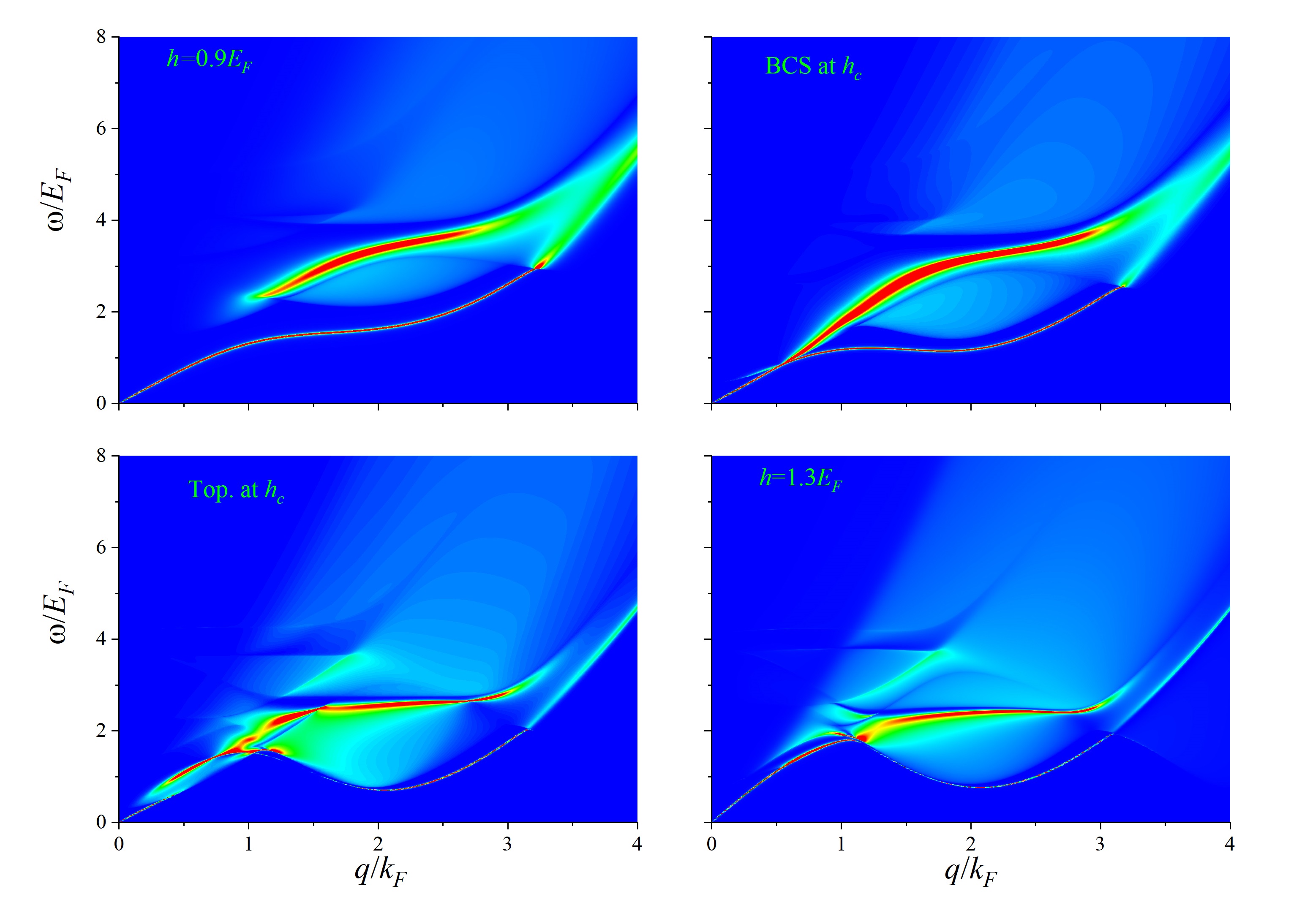}\caption{\label{Fig_dsf_den} The density dynamic structure factor $S_n(q,\omega)$ at different
Zeeman field $h=0.9E_{F},h_{c},1.3E_{F}$.}
\end{figure}

\begin{figure}
\includegraphics[scale=0.35]{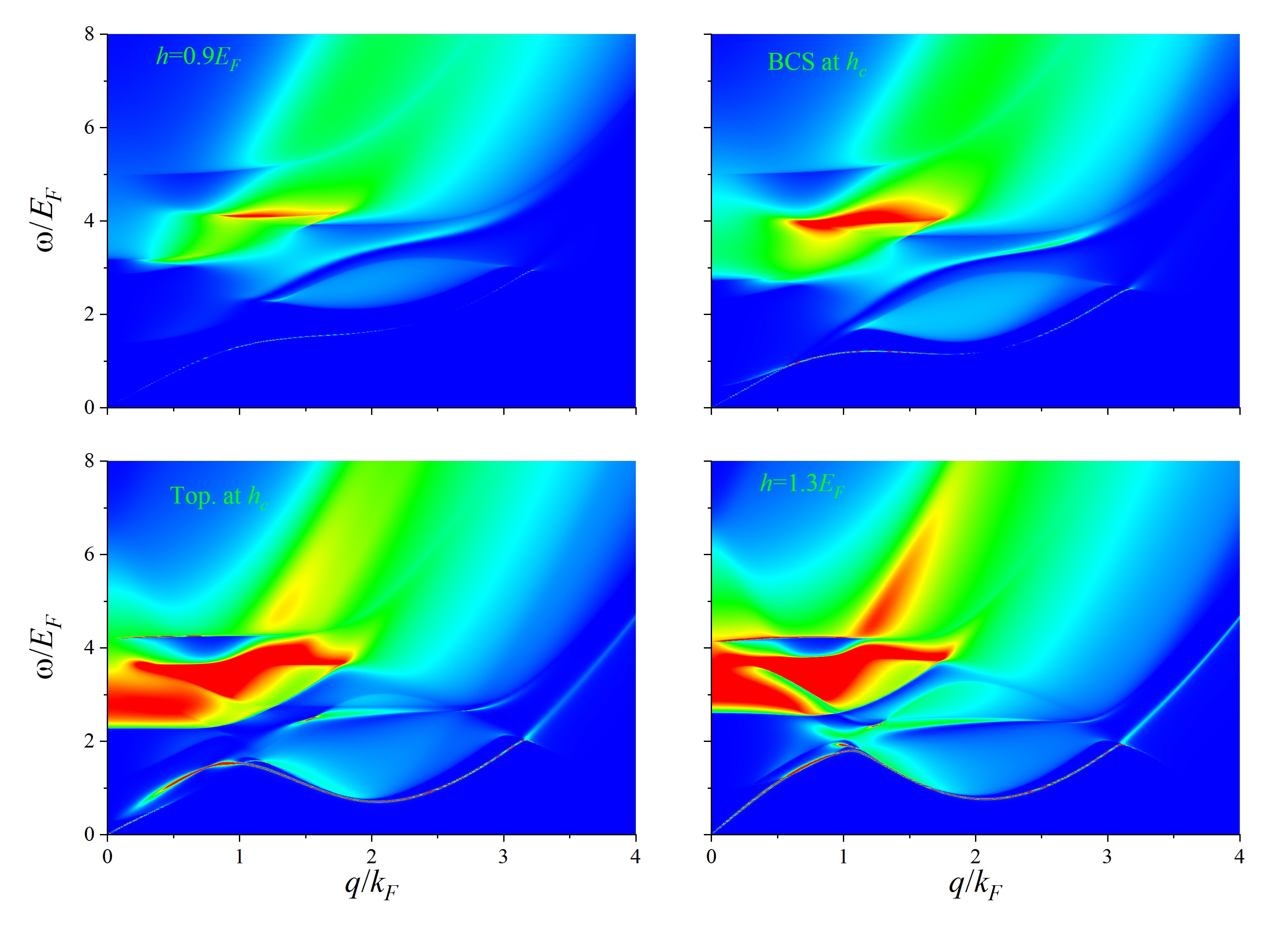}\caption{\label{Fig_dsf_spin} The spin dynamic structure factor $S_s(q,\omega)$ at different
Zeeman field $h=0.9E_{F},h_{c},1.3E_{F}$.}
\end{figure}

In the following discussions, we still focus on the interaction strength
$\gamma=\pi$ and also the recoil momentum $k_{R}=0.75k_{F}$
at zero temperature. These parameters
are the same as one in Fig. \ref{Fig_transition}. We numerically
calculate the density and spin dynamic structure factor, which are shown
in Fig. \ref{Fig_dsf_den} and Fig. \ref{Fig_dsf_spin}, respectively,
in the phase transition between BCS superfluid (higher two panels)
and topological superfluid (lower two panels). Generally we investigate
a full dynamical excitation in different transferred momentum $q$,
including the low energy (or momentum) collective excitation to the
high energy (or momentum) single-particle excitation. Of course, the
presence of SOC effect goes on enriching dynamical behaviors than
the one in conventional Fermi superfluid. 

\subsection{Collective phonon and roton-like excitations}

\begin{figure}
\includegraphics[scale=0.32]{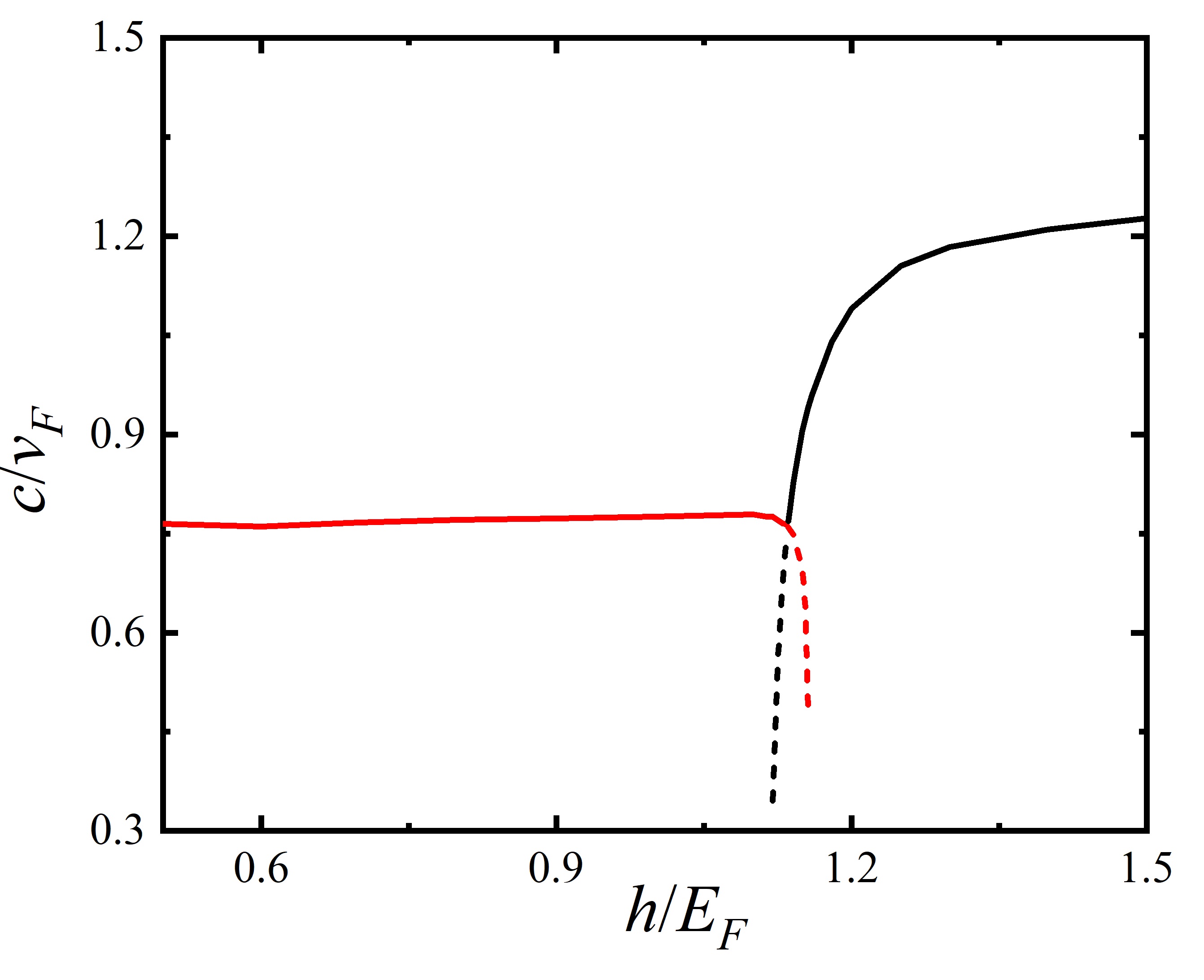}\caption{\label{Fig_sound} The sound velocity $c$ at different effective
Zeeman field $h$.}
\end{figure}

At a low transferred energy $\omega$, it is easy to investigate the
collective excitation. By continuously increasing transferred momentum
$q$ from zero, we initially see a gapless phonon excitation in the
density dynamic structure factor $S_n(q,\omega)$, which is shown by the lower red
curve in all four panels of Fig. \ref{Fig_dsf_den}. When the system is in the BCS superfluid
($h\leq h_{c}$, higher two panels), the spectrum of collective phonon excitation just
monotonically rises with transferred momentum $q$, and finally merges
into the single-particle excitation continuum in a certain large enough $q$. In the whole BCS regime, the
phonon velocity almost does not vary too much with the effective Zeeman field
$h$, except a narrow regime close to transition where BCS superfluid becomes a metastable
state and the velocity suddenly drops (red solid and dot line of Fig. \ref{Fig_sound}).
Of course, the gapless phonon excitation can also be seen in the topological
superfluid (black solid and dot line of Fig. \ref{Fig_sound}), and its velocity monotonically increases
with Zeeman field $h$ and finally saturates to a constant value. In the critical
regime $h=h_{c}$, the BCS and topological state have the same Free energy.
Although we calculate respectively their dynamic structure factor,
the phonon excitation of one state may be potentially influenced by
the other, and turns out a complex excitation behavior (see Fig. \ref{Fig_dsf_den}
and \ref{Fig_dsf_spin}). 

Besides the phonon collective excitation, we investigate a new collective
roton-like excitation only appearing in the topological superfluid. As shown in the lower two panels of both Fig. \ref{Fig_dsf_den} and \ref{Fig_dsf_spin}, this roton-like
excitation is a natural extension of the phonon mode, and it is denoted by a local minimum of the excitation spectrum at a fixed
momentum $q\simeq2k_{F}$, which is just the global minimum of the
down-branch spectrum $D_{k}$ (see red line of Fig. \ref{Fig_uk_spectrum}).
There is no roton-like excitation in the BCS superfluid, where $q\simeq2k_{F}$
is just a local minimum and the global minimum is located at $q=0$.
These results tell us that the emergence of roton-like excitation
is closely related to the formation of global minimum at $q\simeq2k_{F}$,
which is just the character of spectrum $D_k$  in topological superfluid. All discussions above hint that the specific single-particle effect brought by the SOC effect plays an important role in the appearance
of the roton-like excitation at a certain interaction strength. For general, we have
also checked that the same roton-like mode can be seen in other different interaction
strength (for example $\gamma=2.5,4.0$) and recoil momentum $k_{R}$,
and the location of the roton-like excitation is always fixed at
$q\simeq2k_{F}$. 

The dynamical behavior of collective mode can be displayed by both the density and spin dynamic structure factor. However a different excitation related to single-particle excitation happens at a relatively large transferred
energy $\omega$ when $q$ is small, whose physical origin will be introduced in the following.

\subsection{Threshold of single-particle excitation spectrum}

When the transferred energy $\omega$ is large enough, a pair-breaking
of Cooper pairs will occur and make pairs be separated into free Fermi
atoms. Indeed a great part of the dynamical excitation in Fig. \ref{Fig_dsf_den}
and \ref{Fig_dsf_spin} is dominated by this pair-breaking effect.
In the density dynamic structure factor $S_{n}$, this effect usually
is much obvious in a relatively large transferred momentum $q>k_{F}$,
where the collective excitation are depressed very much. Different
from the conventional Fermi superfluid, this single-particle excitation
takes up a large regime in the spin dynamic structure factor $S_{s}$,
even for a small and zero transferred momentum $q$. Before understanding
this single-particle excitation, it is necessary to study the threshold
energy to break a Cooper pair.

\begin{figure}
\includegraphics[scale=0.35]{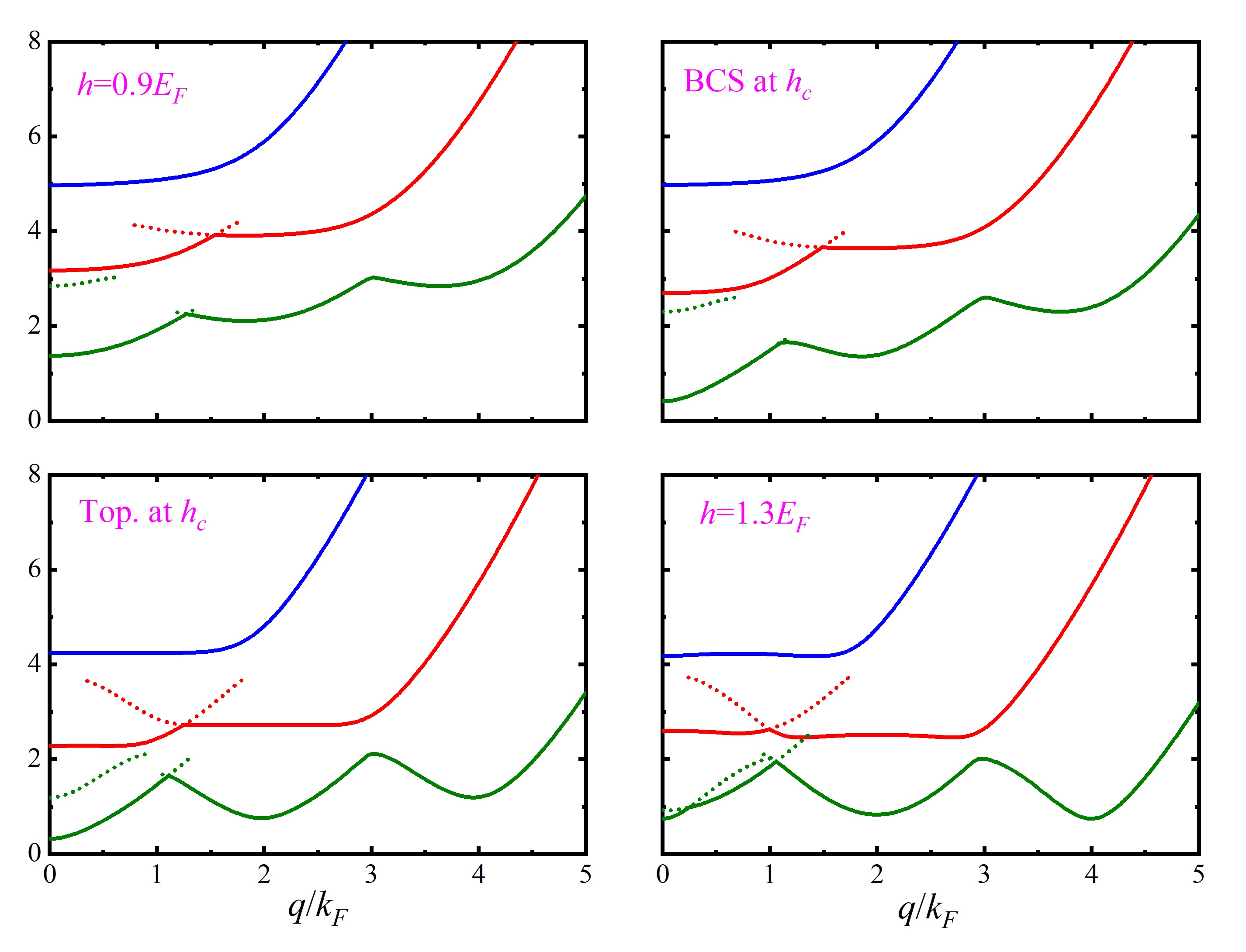}\caption{\label{Fig_spectrum} Four kinds of single-particle excitation spectra.
Olive line: $D_{k}\leftrightarrow D_{k+q}.$ Red line: $D_{k}\leftrightarrow U_{k+q}$
and $U_{k}\leftrightarrow D_{k+q}$. Blue line: $U_{k}\leftrightarrow U_{k+q}$.}
\end{figure}

This pair-breaking excitation is related to two-branch structure of quasi-particle
spectrum $U_{k}$ and $D_{k}$, and the two atoms forming a Cooper-pair
can come from the same or different single-particle spectrum.
This two-branch structure of spectrum generates much richer single-particle excitation than the conventional Fermi superfluid, and induces
four possible combinations of Fermi atoms in a Cooper pair, namely
the $DD$, $DU$, $UD$, and $UU$ type. The minimum energy at a certain
momentum $q$ to break a pair should be ${\rm min}[D_{k}+D_{k+q}]$, ${\rm min}[D_{k}+U_{k+q}]$,
${\rm min}[U_{k}+D_{k+q}]$ or ${\rm min}[U_{k}+U_{k+q}]$. Also due
to the potential three wells geometry of down-branch spectrum $D_{k}$,
there are not only the global minimum energy but also many possible
local minima to break a Cooper pair in single-particle excitations.
These results are shown in Fig. \ref{Fig_spectrum}. The lowest olive
line denote the $DD$ excitation, and its minimum value of pair-breaking
excitation is from the down-branch quasi-particle spectrum (${\rm min}[D_{k}+D_{k+q}]$).
Besides global minimum, it also has two and even three local minima
at some specific transferred momentum $q$, displayed by olive dotted
lines. In other different regime of $q$, these local minima will disappear
since the geometry of spectrum has been changed. From the BCS superfluid
($h=0.9E_{F}$) to the topological superfluid ($h=1.3E_{F}$), the
value of order parameter $\Delta$ monotonically decreases with effective
Zeeman field $h$ (shown by panel (b) of Fig.\ref{Fig_transition}).
This behaviour make the pair-breaking excitation be much easier
in a large $h$, and generally make olive line become lower and lower.

The red line denotes $DU$
and $UD$ excitations, which are overlapped with each other. The two
atoms in a pair comes from different branch of spectrum. It starts
from the ${\rm min}[D_{k}+U_{k+q}]$, whose energy is higher than
the $DD$ one. Similar to $DD$ excitation, there are some possible local
minima in these cross excitations. It should be emphasized that this
$DU$ single-particle excitation at a small $q$ displays a much stronger 
excitation strength in the spin dynamic structure factor than one
in density dynamic structure factor (see Fig. \ref{Fig_dsf_spin}), which also reflect the coupling between different spin components. 

Starting from the ${\rm min}[U_{k}+U_{k+q}]$,
the blue line is the $UU$ excitation, which requires the largest
excitation energy. This excitation has less density of state in the small $q$
regime in the BCS superfluid, while topological state enhances its
density of state and displays a relatively stronger signal. 

All of these kinds of critical single-particle excitations are just
located in the colorful edge curve of Fig.\ref{Fig_dsf_den} and Fig.\ref{Fig_dsf_spin},
and mark the regime of single-particle excitation. To better understand
the dynamical excitation in these colorful panels, we will discuss
the dynamic structure factor at a selected transferred momentum $q$.

\subsection{Dynamic excitation at a constant momentum $q$}

\begin{figure}
\includegraphics[scale=0.35]{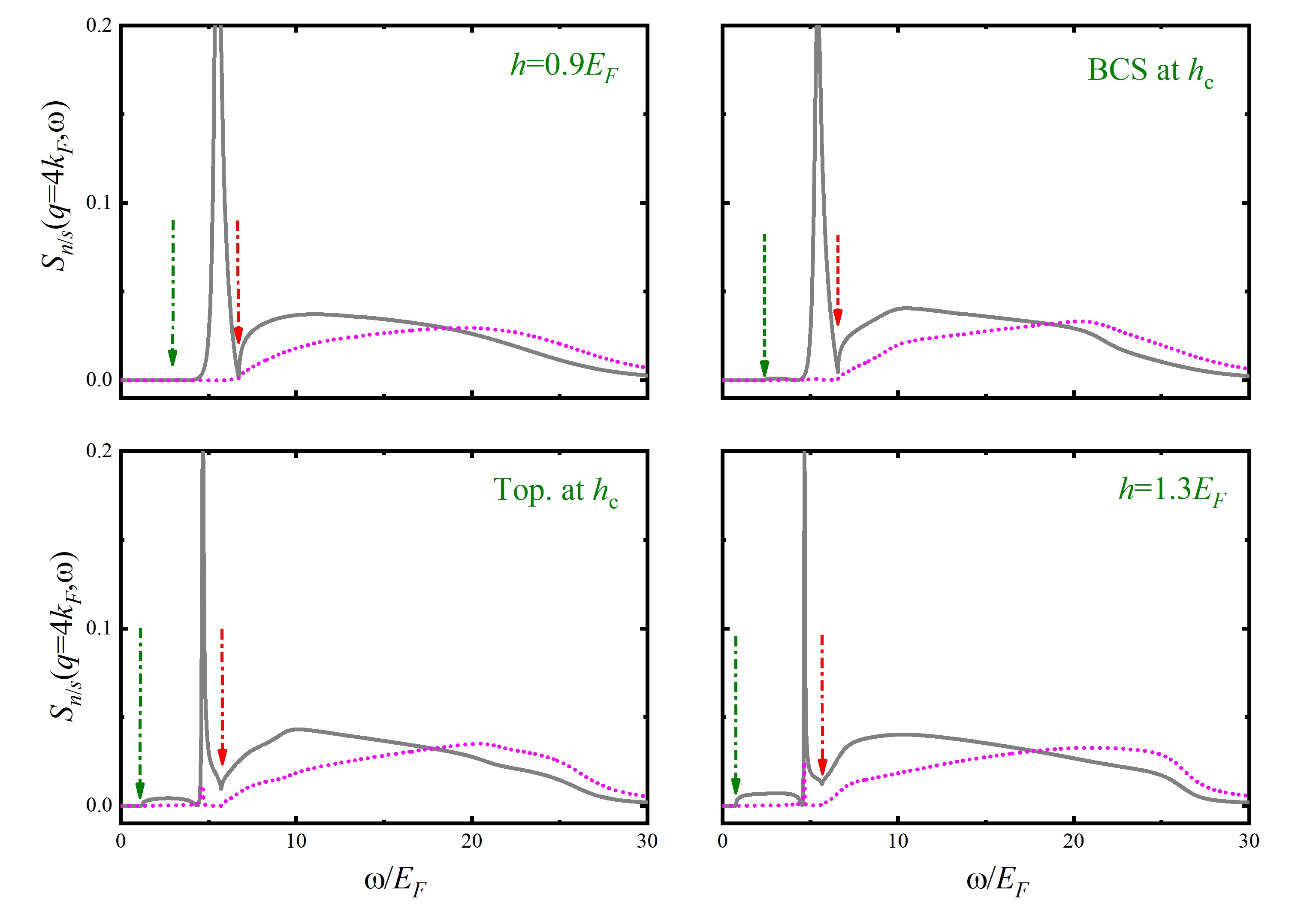}\caption{\label{Fig_q4} The density (gray) and spin (magenta) dynamical structure
factor of 1D SOC Fermi superfluid at transferred momentum $q=4k_{F}$.}
\end{figure}

For a relatively large transferred momentum $q\gg k_{F}$, the dynamic
structure factor will be dominated by the single-particle excitation.
As shown in Fig. \ref{Fig_q4}, we investigate the density and spin
dynamic structure factor at $q=4k_{F}$ in both BCS and topological
superfluid. In all four panels, we can see two obvious single-particle
excitations ($DD$ and $DU$ type) and a sharp collective phonon excitation.
The locations of threshold energy for two single-particle excitations
are respectively labeled by the olive and red dash-dot arrows. Here
the phonon excitation has already been mixed with the $DD$ single-particle excitation, which induces a non-zero expansion width to the
peak of collective mode. Its location is between the olive and red
arrows, after which more and more single-particle excitations turn
out. There is no obvious $UU$ excitation signal here, which is drowned
into the background of other single-particle excitations. At $q=4k_F$, it is easy to see that the topological superfluid displays
a relatively stronger $DD$ excitation $q=4k_{F}$ than one in BCS state. 

\begin{figure}
\includegraphics[scale=0.35]{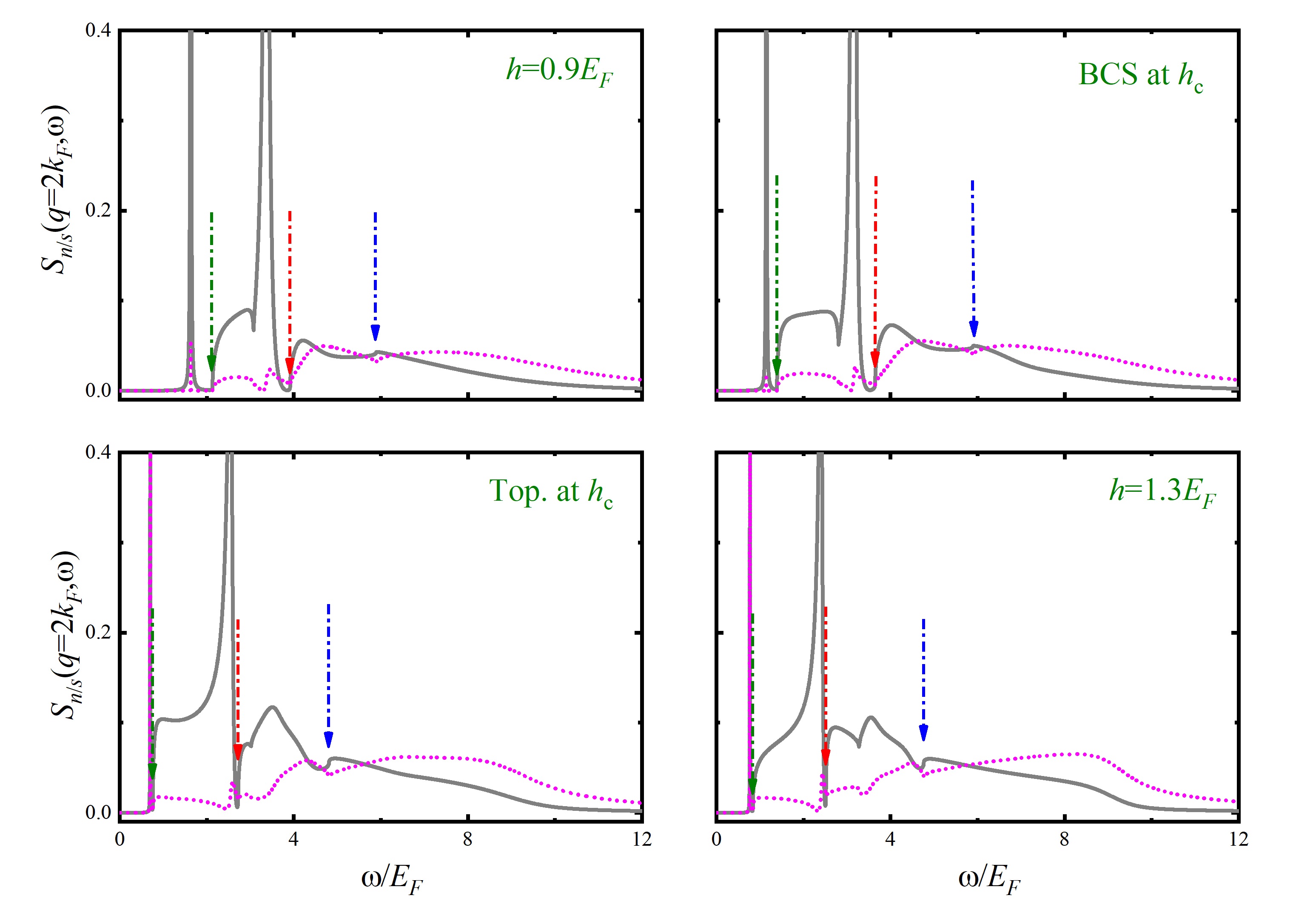}

\caption{\label{Fig_q2} The density (gray) and spin (magenta) dynamical structure
factor of 1D SOC Fermi superfluid at transferred momentum $q=2k_{F}$.}
\end{figure}

When taking transferred momentum $q=2k_{F}$, the competition between
collective mode and single-particle mode is very intense. We watch
a much richer dynamical excitation. As shown in Fig. \ref{Fig_q2},
both $S_{n}$ and $S_{s}$ present two sharp delta-like peak and all
three kinds of single-particle excitation, the threshold locations
of which are still respectively labeled by olive, red and blue arrows.
At $q=2k_{F}$, the left sharp peak in four panels locates on the
left side of green arrows ($DD$ type excitation). In fact it is a
natural extension of the collective phonon excitation, which is close
to merge into the single-particle excitation continuum. The peak in topological
superfluid (lower two panels) happens at a relative low excitation
energy since the system generates a roton-like collective excitation,
which has been discussed above. As to the right sharp peak, it locates at the higher red eyebrow position in Fig. \ref{Fig_dsf_den}, its physical origin is still an open question, we argue that it may be the possible
collective Higgs oscillation, which has totally merged into the single-particle
excitation \citep{Fan2022p}.

\begin{figure}
\includegraphics[scale=0.35]{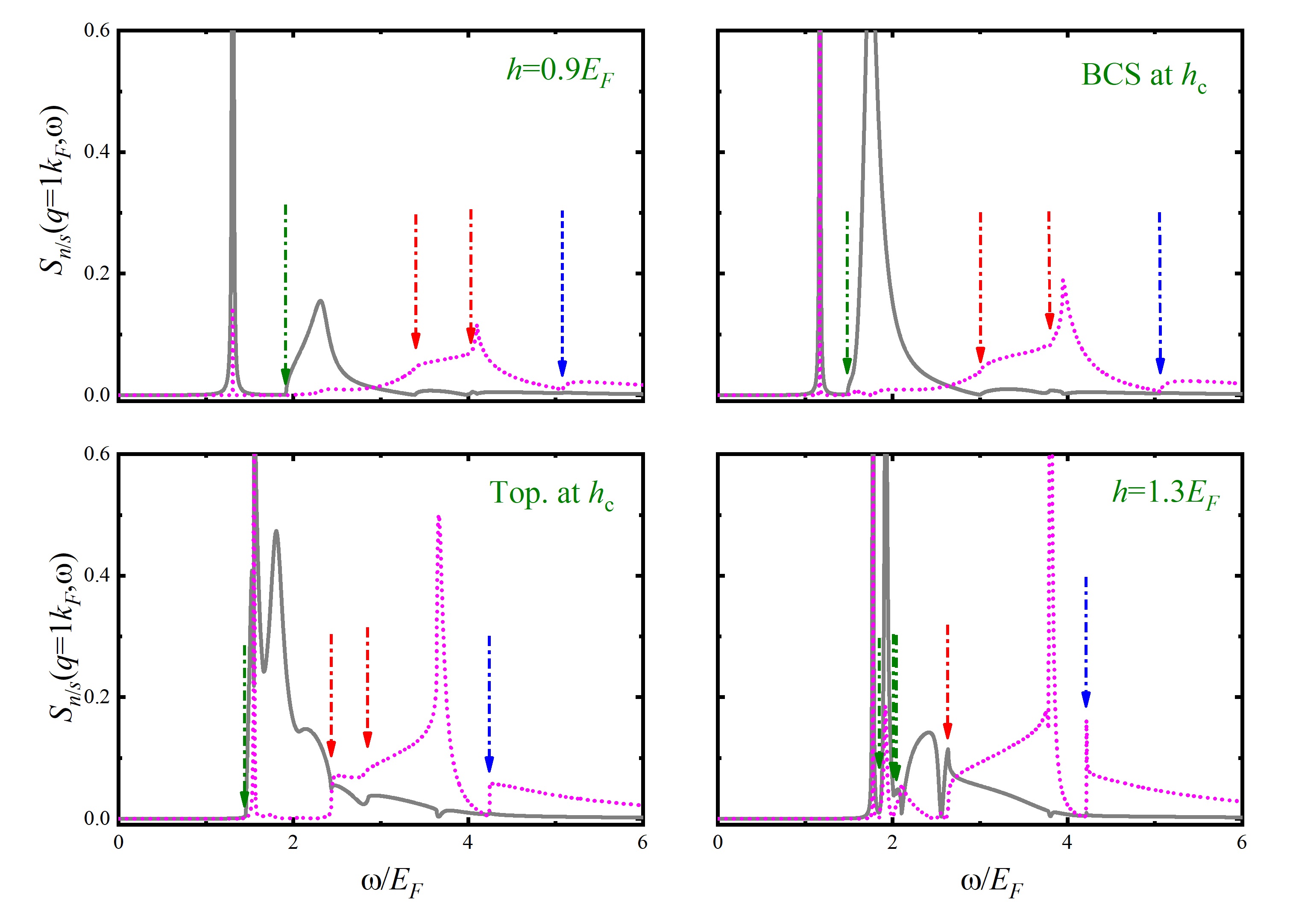}\caption{\label{Fig_q1} The density (blue) and spin (red) dynamical structure
factor of 1D SOC Fermi superfluid at transferred momentum $q=1k_{F}$.}
\end{figure}

For a much smaller transferred momentum $q=1k_{F}$, the competition
of all dynamical excitation displayed by two dynamic structure factors
becomes much more intense, and the energy differences between all possible
excitations are not far away from each other. The results of both BCS and topological state
is shown in Fig. \ref{Fig_q1}. When $h=0.9E_{F}$ (BCS), we see one
clear phonon excitation around $\omega\simeq1.2E_{F}$, and all other
three kinds of single-particle excitations on its right hand, whose initial excitation
energy are still marked by arrows. In this case, $DU$ type excitation
has two threshold energys. While the left red arrow is from the global
minimum of excitation energy ${\rm min}[D_{k}+U_{k+q}]$, the right
one comes from its local minimum. Similar physics also be found in
$h=h_{c}$ (BCS side). However a high peak ($\omega\simeq2.3E_F$) rises after olive arrows. When the system
comes into the topological regime ($h=1.3E_{F}$), this unknown peak ($\omega\simeq1.9E_F$)
will present a delta-like excitation in a certain narrow energy regime
(see also low-right panel of Fig. \ref{Fig_dsf_den}). It seems that this unknown peak is different from the unknown one discussed above. Maybe it is generated by the competition of two collective mode in two different states, and we argue it is the redundancy of collective mode in the metastable state. As to the single
particle excitation, three minima of $DD$ type have also been obtained
in this case, and their positions are located by three olive arrows,
each of which will induce the regular oscillation of the curve of
dynamic structure factor.

\section{Conclusions and outlook}

In summary, we numerically calculate the density and spin dynamic
structure factor of 1D Raman-SOC Fermi superfluid with random phase
approximation during the phase transition between BCS and topological
superfluid. The dynamic structure factor presents rich single-particle
excitations and collective mode. Due to the two-branch structure of
single-particle spectrum, there are three kinds of single-particle
excitation, namely $DD$, $DU$ ($UD$) and $UU$ excitation. We also
calculate their own threshold energy to break a Cooper pair. Among these
single-particle excitation, the $DU$ one takes a great part only
in the spin dynamic structure factor at a small transferred momentum,
which comes from the coupling effect between spin and orbital motion.
As to collective excitation, these is an interesting roton-like collective
excitation at $k\simeq2k_{F}$ when the system comes into the topological
state. The generation of this roton-like excitation is due to the
switch of global minimum of single-particle spectrum $D_{k}$ from
$k=0$ to $k\simeq2k_{F}$. The similar physics has also been found
in other different interaction strengths $\gamma$ and recoil momenta
$k_{R}$. Also these are some unknown quasi-delta like excitations when
$q$ is between $k_{F}$ and $2k_{F}$, which are worth explaining its physical origin in
our future research.

\section{Acknowledgements}

We are grateful for fruitful discussions with Hui Hu, Wei Yi and Wei
Zhang. This research was supported by the National Natural Science
Foundation of China, Grants No. 11804177 (P.Z.), No. 11547034 (H.Z.),
No. 11974384 (S.-G.P.).

\section{Appendix}

In this appendix, we will list expressions of 6 independent Green'
functions and mean-field response function $\chi^{0}=A+B$.

$G_{1}\left(k,\omega\right)=\sum_{l}\left[G_{1}\right]_{k}^{l}/\left(\omega-E_{k}^{l}\right),$
with 

\[
\begin{array}{c}
\left[G_{1}\right]_{k}^{1}=+\frac{U_{k}^{2}-\xi_{-}^{2}-h^{2}-\Delta^{2}}{2\left(U_{k}^{2}-D_{k}^{2}\right)}+\frac{\xi_{+}U_{k}^{2}-\xi_{+}\xi_{-}^{2}+\xi_{-}h^{2}-\xi_{+}\Delta^{2}}{2U_{k}\left(U_{k}^{2}-D_{k}^{2}\right)},\\
\left[G_{1}\right]_{k}^{2}=+\frac{U_{k}^{2}-\xi_{-}^{2}-h^{2}-\Delta^{2}}{2\left(U_{k}^{2}-D_{k}^{2}\right)}-\frac{\xi_{+}U_{k}^{2}-\xi_{+}\xi_{-}^{2}+\xi_{-}h^{2}-\xi_{+}\Delta^{2}}{2U_{k}\left(U_{k}^{2}-D_{k}^{2}\right)},\\
\left[G_{1}\right]_{k}^{3}=-\frac{D_{k}^{2}-\xi_{-}^{2}-h^{2}-\Delta^{2}}{2\left(U_{k}^{2}-D_{k}^{2}\right)}-\frac{\xi_{+}D_{k}^{2}-\xi_{+}\xi_{-}^{2}+\xi_{-}h^{2}-\xi_{+}\Delta^{2}}{2D_{k}\left(U_{k}^{2}-D_{k}^{2}\right)},\\
\left[G_{1}\right]_{k}^{4}=-\frac{D_{k}^{2}-\xi_{-}^{2}-h^{2}-\Delta^{2}}{2\left(U_{k}^{2}-D_{k}^{2}\right)}+\frac{\xi_{+}D_{k}^{2}-\xi_{+}\xi_{-}^{2}+\xi_{-}h^{2}-\xi_{+}\Delta^{2}}{2D_{k}\left(U_{k}^{2}-D_{k}^{2}\right)},
\end{array}
\]
where $\xi_{\pm}=\left(k\pm k_{R}\right)^{2}/2m-\mu$. $G_{2}\left(k,\omega\right)=\sum_{l}\left[G_{2}\right]_{k}^{l}/\left(\omega-E_{k}^{l}\right),$
with

\[
\begin{array}{c}
\left[G_{2}\right]_{k}^{1}=+\frac{U_{k}^{2}-\xi_{+}^{2}-h^{2}-\Delta^{2}}{2\left(U_{k}^{2}-D_{k}^{2}\right)}+\frac{\xi_{-}U_{k}^{2}-\xi_{-}\xi_{+}^{2}+\xi_{+}h^{2}-\xi_{-}\Delta^{2}}{2U_{k}\left(U_{k}^{2}-D_{k}^{2}\right)},\\
\left[G_{2}\right]_{k}^{2}=+\frac{U_{k}^{2}-\xi_{+}^{2}-h^{2}-\Delta^{2}}{2\left(U_{k}^{2}-D_{k}^{2}\right)}-\frac{\xi_{-}U_{k}^{2}-\xi_{-}\xi_{+}^{2}+\xi_{+}h^{2}-\xi_{-}\Delta^{2}}{2U_{k}\left(U_{k}^{2}-D_{k}^{2}\right)},\\
\left[G_{2}\right]_{k}^{3}=-\frac{D_{k}^{2}-\xi_{+}^{2}-h^{2}-\Delta^{2}}{2\left(U_{k}^{2}-D_{k}^{2}\right)}-\frac{\xi_{-}D_{k}^{2}-\xi_{-}\xi_{+}^{2}+\xi_{+}h^{2}-\xi_{-}\Delta^{2}}{2D_{k}\left(U_{k}^{2}-D_{k}^{2}\right)},\\
\left[G_{2}\right]_{k}^{4}=-\frac{D_{k}^{2}-\xi_{+}^{2}-h^{2}-\Delta^{2}}{2\left(U_{k}^{2}-D_{k}^{2}\right)}+\frac{\xi_{-}D_{k}^{2}-\xi_{-}\xi_{+}^{2}+\xi_{+}h^{2}-\xi_{-}\Delta^{2}}{2D_{k}\left(U_{k}^{2}-D_{k}^{2}\right)}.
\end{array}
\]
$\varGamma\left(k,\omega\right)=\sum_{l}\left[\varGamma\right]_{k}^{l}/\left(\omega-E_{k}^{l}\right),$
with

\[
\begin{array}{c}
\left[\varGamma\right]_{k}^{1}=-\left[\varGamma\right]_{k}^{2}=-\frac{\Delta\left[U_{k}^{2}-\left(\xi_{-}^{2}-h^{2}+\Delta^{2}\right)\right]}{2U_{k}\left(U_{k}^{2}-D_{k}^{2}\right)},\\
\left[\varGamma\right]_{k}^{3}=-\left[\varGamma\right]_{k}^{4}=+\frac{\Delta\left[D_{k}^{2}-\left(\xi_{-}^{2}-h^{2}+\Delta^{2}\right)\right]}{2D_{k}\left(U_{k}^{2}-D_{k}^{2}\right)}.
\end{array}
\]
$S\left(k,\omega\right)=\sum_{l}\left[S\right]_{k}^{l}/\left(\omega-E_{k}^{l}\right),$with

\[
\begin{array}{c}
\left[S\right]_{k}^{1}=h\left[-\frac{\xi_{+}+\xi_{-}}{2\left(U_{k}^{2}-D_{k}^{2}\right)}-\frac{U_{k}^{2}+\xi_{+}\xi_{-}-h^{2}+\Delta^{2}}{2U_{k}\left(U_{k}^{2}-D_{k}^{2}\right)}\right],\\
\left[S\right]_{k}^{2}=h\left[-\frac{\xi_{+}+\xi_{-}}{2\left(U_{k}^{2}-D_{k}^{2}\right)}+\frac{U_{k}^{2}+\xi_{+}\xi_{-}-h^{2}+\Delta^{2}}{2U_{k}\left(U_{k}^{2}-D_{k}^{2}\right)}\right],\\
\left[S\right]_{k}^{3}=h\left[+\frac{\xi_{+}+\xi_{-}}{2\left(U_{k}^{2}-D_{k}^{2}\right)}+\frac{D_{k}^{2}+\xi_{+}\xi_{-}-h^{2}+\Delta^{2}}{2D_{k}\left(U_{k}^{2}-D_{k}^{2}\right)}\right],\\
\left[S\right]_{k}^{4}=h\left[+\frac{\xi_{+}+\xi_{-}}{2\left(U_{k}^{2}-D_{k}^{2}\right)}-\frac{D_{k}^{2}+\xi_{+}\xi_{-}-h^{2}+\Delta^{2}}{2D_{k}\left(U_{k}^{2}-D_{k}^{2}\right)}\right].
\end{array}
\]
$F_{1}\left(k,\omega\right)=\sum_{l}\left[F_{1}\right]_{k}^{l}/\left(\omega-E_{k}^{l}\right),$
with

\[
\begin{array}{c}
\left[F_{1}\right]_{k}^{1}=-\frac{\Delta h\left(2U_{k}+\xi_{+}-\xi_{-}\right)}{2U_{k}\left(U_{k}^{2}-D_{k}^{2}\right)},\\
\left[F_{1}\right]_{k}^{2}=-\frac{\Delta h\left(2U_{k}-\xi_{+}+\xi_{-}\right)}{2U_{k}\left(U_{k}^{2}-D_{k}^{2}\right)},\\
\left[F_{1}\right]_{k}^{3}=+\frac{\Delta h\left(2D_{k}+\xi_{+}-\xi_{-}\right)}{2D_{k}\left(U_{k}^{2}-D_{k}^{2}\right)},\\
\left[F_{1}\right]_{k}^{4}=+\frac{\Delta h\left(2D_{k}-\xi_{+}+\xi_{-}\right)}{2D_{k}\left(U_{k}^{2}-D_{k}^{2}\right)}.
\end{array}
\]
$F_{2}\left(k,\omega\right)=\sum_{l}\left[F_{2}\right]_{k}^{l}/\left(\omega-E_{k}^{l}\right),$
with

\[
\begin{array}{c}
\left[F_{2}\right]_{k}^{1}=+\frac{\Delta h\left(2U_{k}-\xi_{+}+\xi_{-}\right)}{2U_{k}\left(U_{k}^{2}-D_{k}^{2}\right)},\\
\left[F_{2}\right]_{k}^{2}=+\frac{\Delta h\left(2U_{k}+\xi_{+}-\xi_{-}\right)}{2U_{k}\left(U_{k}^{2}-D_{k}^{2}\right)},\\
\left[F_{2}\right]_{k}^{3}=-\frac{\Delta h\left(2D_{k}-\xi_{+}+\xi_{-}\right)}{2D_{k}\left(U_{k}^{2}-D_{k}^{2}\right)},\\
\left[F_{2}\right]_{k}^{4}=-\frac{\Delta h\left(2D_{k}+\xi_{+}-\xi_{-}\right)}{2D_{k}\left(U_{k}^{2}-D_{k}^{2}\right)}.
\end{array}
\]

The expressions of all 9 independent matrix elements in mean-field
response function $A$ are

$A_{11}=+\underset{pll'}{\sum}\left[G_{1}\right]_{p}^{l}\left[G_{1}\right]_{p+q}^{l'}\frac{f\left(E_{p}^{l}\right)-f\left(E_{p+q}^{l'}\right)}{i\omega_{n}+E_{p}^{l}-E_{p+q}^{l'}},$

$A_{12}=-\underset{pll'}{\sum}\left[\varGamma\right]_{p}^{l}\left[\varGamma\right]_{p+q}^{l'}\frac{f\left(E_{p}^{l}\right)-f\left(E_{p+q}^{l'}\right)}{i\omega_{n}+E_{p}^{l}-E_{p+q}^{l'}},$

$A_{13}=+\underset{pll'}{\sum}\left[G_{1}\right]_{p}^{l}\left[\varGamma\right]_{p+q}^{l'}\frac{f\left(E_{p}^{l}\right)-f\left(E_{p+q}^{l'}\right)}{i\omega_{n}+E_{p}^{l}-E_{p+q}^{l'}},$

$A_{14}=+\underset{pll'}{\sum}\left[\varGamma\right]_{p}^{l}\left[G_{1}\right]_{p+q}^{l'}\frac{f\left(E_{p}^{l}\right)-f\left(E_{p+q}^{l'}\right)}{i\omega_{n}+E_{p}^{l}-E_{p+q}^{l'}},$

$A_{22}=+\underset{pll'}{\sum}\left[G_{2}\right]_{p}^{l}\left[G_{2}\right]_{p+q}^{l'}\frac{f\left(E_{p}^{l}\right)-f\left(E_{p+q}^{l'}\right)}{i\omega_{n}+E_{p}^{l}-E_{p+q}^{l'}},$

$A_{23}=-\underset{pll'}{\sum}\left[\varGamma\right]_{p}^{l}\left[G_{1}\right]_{p+q}^{l'}\frac{f\left(E_{p}^{l}\right)-f\left(E_{p+q}^{l'}\right)}{i\omega_{n}+E_{p}^{l}-E_{p+q}^{l'}},$

$A_{24}=-\underset{pll'}{\sum}\left[G_{1}\right]_{p}^{-l}\left[\varGamma\right]_{p+q}^{l'}\frac{f\left(E_{p}^{l}\right)-f\left(E_{p+q}^{l'}\right)}{i\omega_{n}+E_{p}^{l}-E_{p+q}^{l'}},$

$A_{34}=+\underset{pll'}{\sum}\left[G_{2}\right]_{p}^{-l}\left[G_{2}\right]_{p+q}^{l'}\frac{f\left(E_{p}^{l}\right)-f\left(E_{p+q}^{l'}\right)}{i\omega_{n}+E_{p}^{l}-E_{p+q}^{l'}},$

$A_{43}=+\underset{pll'}{\sum}\left[G_{1}\right]_{p}^{l}\left[G_{1}\right]_{p+q}^{-l'}\frac{f\left(E_{p}^{l}\right)-f\left(E_{p+q}^{l'}\right)}{i\omega_{n}+E_{p}^{l}-E_{p+q}^{l'}},$

where $f\left(x\right)=1/\left(e^{x/k_{B}T}+1\right)$ is the Fermi-Dirac
distribution function. The expressions of 10 independent matrix elements
in mean-field response function $B$ are

$B_{11}=-\underset{pll'}{\sum}\left[F_{1}\right]_{p}^{l}\left[F_{1}\right]_{p+q}^{l'}\frac{f\left(E_{p}^{l}\right)-f\left(E_{p+q}^{l'}\right)}{i\omega_{n}+E_{p}^{l}-E_{p+q}^{l'}},$

$B_{12}=+\underset{pll'}{\sum}\left[S\right]_{p}^{l}\left[S\right]_{p+q}^{l'}\frac{f\left(E_{p}^{l}\right)-f\left(E_{p+q}^{l'}\right)}{i\omega_{n}+E_{p}^{l}-E_{p+q}^{l'}},$

$B_{13}=-\underset{pll'}{\sum}\left[S\right]_{p}^{l}\left[F_{1}\right]_{p+q}^{l'}\frac{f\left(E_{p}^{l}\right)-f\left(E_{p+q}^{l'}\right)}{i\omega_{n}+E_{p}^{l}-E_{p+q}^{l'}},$

$B_{14}=-\underset{pll'}{\sum}\left[F_{1}\right]_{p}^{l}\left[S\right]_{p+q}^{l'}\frac{f\left(E_{p}^{l}\right)-f\left(E_{p+q}^{l'}\right)}{i\omega_{n}+E_{p}^{l}-E_{p+q}^{l'}},$

$B_{22}=-\underset{pll'}{\sum}\left[F_{2}\right]_{p}^{l}\left[F_{2}\right]_{p+q}^{l'}\frac{f\left(E_{p}^{l}\right)-f\left(E_{p+q}^{l'}\right)}{i\omega_{n}+E_{p}^{l}-E_{p+q}^{l'}},$

$B_{23}=+\underset{pll'}{\sum}\left[S\right]_{p}^{l}\left[F_{2}\right]_{p+q}^{l'}\frac{f\left(E_{p}^{l}\right)-f\left(E_{p+q}^{l'}\right)}{i\omega_{n}+E_{p}^{l}-E_{p+q}^{l'}},$

$B_{24}=+\underset{pll'}{\sum}\left[F_{2}\right]_{p}^{l}\left[S\right]_{p+q}^{l'}\frac{f\left(E_{p}^{l}\right)-f\left(E_{p+q}^{l'}\right)}{i\omega_{n}+E_{p}^{l}-E_{p+q}^{l'}},$

$B_{33}=-\underset{pll'}{\sum}\left[F_{2}\right]_{p}^{l}\left[F_{1}\right]_{p+q}^{l'}\frac{f\left(E_{p}^{l}\right)-f\left(E_{p+q}^{l'}\right)}{i\omega_{n}+E_{p}^{l}-E_{p+q}^{l'}},$

$B_{34}=-\underset{pll'}{\sum}\left[S\right]_{p}^{-l}\left[S\right]_{p+q}^{l'}\frac{f\left(E_{p}^{l}\right)-f\left(E_{p+q}^{l'}\right)}{i\omega_{n}+E_{p}^{l}-E_{p+q}^{l'}},$

$B_{43}=-\underset{pll'}{\sum}\left[S\right]_{p}^{l}\left[S\right]_{p+q}^{-l'}\frac{f\left(E_{p}^{l}\right)-f\left(E_{p+q}^{l'}\right)}{i\omega_{n}+E_{p}^{l}-E_{p+q}^{l'}}.$

\end{document}